\documentclass[superscriptaddress]{revtex4}
\usepackage{amsmath}
\usepackage{hyperref}
\usepackage{color}
\usepackage{latexsym}

\newcommand{\im}{{\mathrm{i}}}

\renewcommand{\and}{{\quad{\rm and}\quad}}

\def\DH{\rm I\kern-1.5pt\rm H\kern-1.5pt\rm I}

\def\DR{\rm I\kern-1.45pt\rm R}
\def\DC{\kern2pt {\hbox{\sqi I}}\kern-4.2pt\rm C}

\newcommand{\cF}{{\cal F}}

\newcommand{\bQ}{{\overline Q}}

\newcommand{\bpsi}{{\bar\psi}}

\newcommand{\ba}{\begin{array}}
\newcommand{\ea}{\end{array}}
\newcommand{\be}{\begin{equation}}
\newcommand{\ee}{\end{equation}}
\newcommand{\bea}{\begin{eqnarray}}
\newcommand{\eea}{\end{eqnarray}}
\newcommand{\bi}{\begin{itemize}}
\newcommand{\ei}{\end{itemize}}

\begin{document}

\title{Euler top and freedom in supersymmetrization of one-dimensional mechanics }

\author{Erik Khastyan}
\email{khastyan.erik@gmail.com}
\affiliation{Yerevan Physics Institute, 2 Alikhanian Brothers St., Yerevan  0036 Armenia}

\author{Sergey Krivonos}
\email{krivonos@theor.jinr.ru}
\affiliation{Bogoliubov Laboratory of Theoretical Physics, Joint Institute for Nuclear Research, Dubna, Russia}

\author{Armen Nersessian}
\email{arnerses@yerphi.am}
\affiliation{Yerevan Physics Institute, 2 Alikhanian Brothers St., Yerevan  0036 Armenia}
\affiliation{Bogoliubov Laboratory of Theoretical Physics, Joint Institute for Nuclear Research, Dubna, Russia}
\affiliation{Institute of Radiophysics and Electronics, Ashtarak-2,0203, Armenia }
\begin{abstract}
Recently A.Galajinsky  has suggested the $\mathcal{N}=1$ supersymmetric extension of Euler top and made a few interesting observations on its properties \cite{AntonEuler}.  In this paper we use the  formulation of the Euler top as a system on complex projective plane,  playing the role of phase space,  i.e.  as a one-dimensional mechanical system.
 Then we suggest the supersymmetrization scheme of the generic one-dimensional systems  with positive Hamiltonian
which yields   {\'a priori} integrable family of $\mathcal{N}=2k$ supersymmetric Hamiltonians parameterized by    $\mathcal{N}/2$ arbitrary real functions.
\end{abstract}

\maketitle
\section{Introduction}
The Euler top  is a textbook integrable system describing the rotation of free rigid body, see, e.g.  \cite{arnold}.
 Conventionally it is described  by the Hamiltonian system  with degenerated Poisson brackets parameterized by the components
 of angular momentum $\ell=(x_1,x_2,x_3)$,
\be
 \{x_i,x_j\}=\varepsilon_{ijk}x_k,\qquad  H=\sum_{i=1}^3 \frac{x^2_i}{2I_i},
\label{euler} \ee
 where $H$ is the Hamiltonian,  and $I_i>0$ are the principal momenta of inertia.
 Since $x_i$ form $so(3)$ algebra, the system has a Casimir function
 \be
 C=\sum_{i=1}^3 x^2_i  \; :\quad \{C(x), x_i\}=0
 \label{casimir}\ee
Its fixation  leads to the Hamiltonian system with two-dimensional {\bf non-degenerated} phase space, i.e.  one-dimensional system.  Hence, Euler top is {\sl a priori }  integrable.
 Being introduced centuries ago,  Euler top has been studied as completely as the one-dimensional  oscillator both at classical and quantum-mechanical levels. So, none of the open questions is to be studied there, except various aspects of its perturbations and generalizations.

Very recently A.Galajinsky noticed the absence of the relevant   supersymmetric extensions of Euler top \cite{AntonEuler}. He
 suggested  its  $\mathcal{N}=1$ supersymmetrization via  extension of
  the degenerated Poisson brackets \eqref{euler} by  {\sl three real } Grassmann coordinates,  stating that in general the resulting system lacks integrability.
It seems to us that   many questions asked in that paper  come  from the improper supersymmetrization procedure  formulated in terms of degenerated phase space. As a consequence, it yields
  overcompleted  number of fermionic  variables which do not have impact on the actual properties of the system.
  Moreover, the invention of three Grassmann variables could yield the problems with the physical interpretation of the quantized version of system (though quantum aspects were not touched in that paper).
  Furthermore,  the $\mathcal{N}\geq 2$ supersymmetric extensions of the Euler top, which at the quantum level create  qualitative corrections to the initial spectrum (e.g.  degeneracy of the energy levels etc), were  not considered there at all.

In this paper we  propose to supersymmetrize the Euler top formulated in terms of nondegenerated phase space.
First, fixing the value of the Casimir \eqref{casimir},  we formulate the Hamiltonian system \eqref{euler} in terms of  two-dimensional nondegenerated phase space  given by the complex projective plane $\mathbf{CP}^1$, i.e. as a   one-dimensional mechanical system. Since any two-dimensional manifold can be provided with the K\"ahler structure, the initial system can be quantized by the well-developed technique of  geometric quantization on K\"ahler manifold (see, e.g. \cite{nair}).

Then we present the   procedure of $\mathcal{N}=2k$ supersymmerization of the systems with generic two-dimensional nondegenerated phase space which  results in {\sl \'a priory} integrable supersymmetric extension of the initial system.
Suggested   procedure provides us with the family of the supersymmetric systems parameterized by the
$\mathcal{N}/2$ angle-like arbitrary functions. Similar functional  freedom   was noticed earlier in the one-dimensional $\mathcal{N}=2$ supersymmetric mechanics \cite{dihedral} and in the two-dimensional $\mathcal{N}=4$ supersymmetric  ones \cite{freedom}.
In proposed supersymmetrization scheme the fermionic variables are splitted from the bosonic ones, in contrast with $\mathcal{N}=2$ supersymmetrization procedure of the systems  with generic K\"ahler phase space   suggested in \cite{npps}. They form, with respect to the Poisson brackets,  the Clifford algebra. Thus,  guantization of the supersymmetric system is   straightforward: we should  perform geometric quantization of the initial bosonic system and then  replace  the fermionic variables by the respective gamma-matrices.\\

The paper  is organized as follows.

In {\sl  Section 2} we review the description  of  Euler top on  the phase space given by the complex projective plane $\mathbf{CP}^1$.

In {\sl Section 3}  we present the $\mathcal{N}=2k $ supersymmetization procedure for the system with generic one-(complex)dimensional K\"ahler spaces.

In {\sl Section 4} we summarize  obtained results and  discuss the possible extensions of the proposed scheme to the Lagrange and Kowalewski tops.
%
%
%

\section{Euler top}
For the description of the Euler top \eqref{euler} in terms of the nondegenarated phase space, let us introduce instead of $x_i$, the  coordinates $j, z, {\bar z}$
\be
j:= \sqrt{\sum_{i=1}^3x^2_i}, \qquad z :=\frac{x_1+\imath x_2}{j-x_3}.
\label{z}
\ee
Clearly, $j$ is the complete angular momentum.

In these coordinates the Poisson brackets read
\be
\{\bar z, z\}=-\frac{\imath}{2j}(1+z \bar z )^2, \quad \{z , j\}=\{z , z \}=0,
\label{pbr}\ee
while the  momentum generators  look    as follows
 \be
j^2:= \sum_{i=1}^3x^2_i,\qquad x_1:=h_1=j\frac{z+\bar z }{1+z\bar z},\qquad  x_2:=h_2=j\frac{\imath(\bar z - z ) }{1+z \bar z }\;,\quad x_3:=h_3=  j \frac{z \bar z  -1}{1+z \bar z }\;.
\label{killing1}\ee
However, the point $x_3=  j$ cannot be described in these terms.
To improve this lack we should introduce, instead of $z$,  another coordinates ${\tilde z},\bar{\tilde z}$
 \be
{\tilde z} :=\frac{x_1-\imath x_2}{j+x_3}\;:\quad \{\bar{\tilde z}, {\tilde z}\}=-\frac{\imath}{2j}(1+{\tilde z} \bar{\tilde  z} )^2.
\label{tz}
\ee
 Out of the points $x_3=\pm j$ these coordinates are related with each other as follows
 \be
 {\tilde z}=\frac 1z\;.
 \ee
The Poisson brackets \eqref{pbr} and \eqref{tz}  transform to each other    under this transformation.
Thus, fixing $j$ to be constant
 we arrive  at the two-dimensional phase space covered by two charts
(parameterized   by the  single complex coordinate $z$ or ${\tilde z}$ )
 and equipped with the
  one-(complex)dimensional K\"ahler structure - the
complex projective plane  $\mathbf{CP}^1$   with the Fubini-Study metrics
\be
g(z,\bar z)dzd\bar z :=2j\frac{dz d\bar z}{(1+z\bar z)^2},
\label{fs}\ee
 which corresponds to the K\"ahler potential
\be
K(z,\bar z)= { 2j}\log(1+z\bar z).
\label{fskah}\ee
The generators \eqref{killing1} become  Killing potentials of $\mathbf{CP^1}$, and  define the   Hamiltonian holomorphic vector fields,
\be
\{h_1,\quad\}=\imath (1-z^2)\partial_z +c.c.,\qquad \{h_2,\quad\}= -(1+z^2)\partial_z +c.c.,\qquad \{h_3,\quad\}= 2\imath z\partial_z +c.c.,
\ee

In these terms the Hamiltonian of Euler top reads
\be
H=\sum_{i=1}^3\frac{x^2_i}{2I_i}=-j^2\frac{ b(z^2+\bar z^2)+ 2az\bar z }{2(1+z\bar z)^2}  +\frac{j^2}{2I_3},
\ee
where
\be
a:= \frac{2}{I_3}-\frac{1}{I_1}-\frac{1}{I_2}, \qquad b:= \frac{1}{I_2}-\frac{1}{I_1}.
\ee

Now, let us rewrite   the Euler top in canonical coordinates.
For this purpose we  notice that  $\mathbf{CP}^1$ is just the two-dimensional sphere $\mathbf{S}^2$ formulated in the projective coordinates
\be
z= \cot\frac{\theta}{2}{\rm e}^{\imath\varphi},
\ee
where $\theta, \varphi$ are the spherical coordinates. In these terms the Poisson bracket \eqref{pbr} reads
\be
\{\varphi,j \cos\theta\}=1
\ee
Hence, the function  $p:=j\cos\theta$ defines  the canonical momentum conjugated to $\varphi$.
In terms of these canonical coordinates the angular momentum generators (and Killing potentials) \eqref{killing1} read
\be
x_1+\imath x_2= j\sin\theta{\rm e}^{\imath\varphi}=  \sqrt{j^2-p^2}{\rm e}^{\imath\varphi} ,\quad x_3= j\cos\theta= p,
\ee
while the Hamiltonian of   the Euler top takes the form
\be\label{HEt}
H=\frac14(a+b \cos 2\varphi)p^2+\frac{j^2}{4}\left(\frac{2}{I_3}-(a+b \cos 2\varphi)\right).
\ee

Without loss of generality we assume
\be
I_3\leq I_2\leq I_1,
\ee
and  perform   canonical transformation
$(p,\varphi)\to (P,Q)$,
where
\be
P=\sqrt{\frac{a+b\cos2\varphi}{2}}\;p,\qquad Q=\sqrt{\frac{2}{a+b}}\int\frac{d\varphi}{\sqrt{1-\frac{2b}{a+b}\sin^2 \varphi}}= \sqrt{\frac{2}{a+b}}F\left(\varphi,k\right):\qquad \{Q,P\}=1,
\ee
where $F\left(\varphi,k\right)$ is an elliptic integral of the first kind, with $k=\sqrt{2b/(a+b)}$ being its modulus, and $\varphi$ is the so-called Jacobi amplitude (which defines the Jacobi elliptic functions) \cite{dwait},
\be
\varphi = F^{-1} \left(F,k\right) = \textbf{amp}\left(F,k\right).
\ee
Respectively
\be
\sin \varphi = \sin (\textbf{amp}(F,k) ) = \textbf{sn}(F,k)
\ee
is the  Jacobi sine 
amplitude of  the elliptic functions.

In this terms the Hamiltonian of Euler top reads
\be
H=\frac12 P^2+\frac{j^2b}{2} \textbf{sn}^2 \left(\sqrt{\frac{a+b}{2}}Q,\sqrt{\frac{2b}{a+b}}\right)
  +\frac{j^2}{2I_1}.
\ee
In the particular case of  the symmetric top ($I_1=I_2:=I$) it  reduces to the one-dimensional free particle Hamiltonian
\be
H=\frac12\left(\frac{1}{I_3}-\frac{1}{I}\right)p^2+\frac{j^2}{2I}.
\ee
{\sl So,  the  Euler top  is the  one-dimensional Hamiltonian system   with   $\mathbf{CP}^1$ phase space and with the Hamiltonian given by the quadratic functions of its Killing potentials. In the canonical coordinates it results in the one-dimensional nonlinear  oscillator.}

\section{Supersymmetry}
In the previous section we formulated  the Euler top in terms of  one-(complex)dimensional phase space given by complex projective plane.
Being one-dimensional system,  the Euler top allows many ways of supersymmetrization, including supersymmetrization  in canonical coordinates.
However, we are interested   in the supersymmetrization compatible with the  K\"ahler geometry describing the phase spase of the Euler top.

One of the ways to supersymmetrize  the Euler top is to use the approach suggested in \cite{npps} which is based  on the extension of the K\"ahler phase space to the super-K\"ahler one
defined by the
potential
\be
\mathcal{K}(z,\bar z, \theta_a,\bar\theta^a)=K(z,\bar z) + F(\imath g(z,\bar z)\theta_a{\bar\theta}^{a} ),
\label{sK}\ee
where $F(x)$ is the real function with $F'(0)\neq 0$,  with $K(z,\bar z)$, $g(z,\bar z)$ given by \eqref{fskah} and \eqref{fs},
while  the fermionic variables $\theta_a$ are associated with $dz$,  in complete similarity with the superfield approach.

Another particular way of supersymmetrization is to extend the complex projective plane to the complex projective super-plane given by
the K\"ahler potential
  \be
  {\tilde K}(z,\bar z,\theta_a,   \bar\theta^a)=2j\log(1+z\bar z+\theta_a\bar\theta^a).
  \ee
Such approach has been taken  in \cite{lobach}, where it was applied to the Lobachevsky plane (i.e. non-compact version of complex projective plane)
for the construction of  $\mathcal{N}$-extended one-dimensional superconformal mechanics.  Later on, this approach was generalized to the higher-dimensional systems in \cite{kkn}.

Below we suggest    different,   less geometric approach, which is applicable   not only for the Euler top, but for any one-dimensional system.    We will consider     the
systems with  generic two-(real)dimensional phase space.  Such phase spaces can be always equipped with the one-(complex)dimensional
 K\"ahel structure, so that  the Poisson brackets will be given by the relation
\be
\{z,\bar z\}=\frac{\imath}{g(z,\bar z)}.
\label{pbg}\ee

For the construction of  $\mathcal{N}$-supersymmetric extensions of these systems ( with even $\mathcal{N}$  )
we extend this phase space by the canonical complex Grassmann variables $\psi_a$, $ a=1,\ldots,\frac{\mathcal{N}}{2}$
\be\label{PB2}
\left\{ \psi_a, \bpsi^b \right\} = \imath \delta_a^b , 
\ee
where  $\overline{\left(\psi_a\right)}:=\bpsi^a$  and
 $\overline{  \cF_1 \cF_2}  = \overline{\cF}_2\overline{\cF_1} $.

 With these Poisson brackets at hands we can construct the $\mathcal{N}$ supersymmetric extensions of two-dimensional systems defined by the Poisson brackets \eqref{pbg} and by any  positive  Hamiltonian  $H(z,\bar z)>0$,
 \be\label{Poincare}
\left\{ Q_a, \bQ^b\right\} = {\imath} \delta_a^b \mathcal{H}\;,\qquad \mathcal{H}:=H(z,\bar z)+\; fermions \quad .
\ee
In accordance with the generalization  of Liouville theorem to the supermanifolds \cite{shander}(see also \cite{dihedral})
these supersymmetric extensions will be {\sl a\' priory} integrable, since we will get the system with $(2|\mathcal{N})_{\mathbf{R}}$-dimensional phase space with  one bosonic constant of motion $\mathcal{H}$ and $\mathcal{N}$ fermionic constants of motion $Q_a,\bQ^b $
commuting with the bosonic integral $\mathcal{H}$.

\subsection*{$\mathcal{N}=2$ supersymmetry}
For the construction of $\mathcal{N}=2$ supersymmetric extension of the system with Hamiltonian $H(z,\bar z)>0$ we choose,
following \cite{dihedral}, the appropriate Ansatz for supercharges and  arrive   the  family of $\mathcal{N}=2$ supersymmetric extensions of the Hamiltonian $H$, parameterized by the  arbitrary real function $\Phi(z,\bar z)$
\be
Q=\sqrt{H}{\rm e}^{\imath\Phi}\psi, \qquad \bQ=\sqrt{H}{\rm e}^{-\imath\Phi}\bpsi\quad \Rightarrow\quad \mathcal{H}=H+\{\Phi,H\}\psi\bpsi.
\ee
 Specifying the Poisson brackets and Hamiltonian we will get the respective supersymmetric extension of the Euler top.

Direct extension of construction    to the $\mathcal{N} \geq 4 $ supersymmetric mechanics fixes the  function $\Phi $ and leads to the trivial family of the supersymmetric Hamiltonians. Namely, choosing
$ Q_a = \sqrt{H}{\rm e}^{\imath\Phi} \psi_a $,
we get that the superalgebra \eqref{Poincare} is fulfilled when $\{H,\Phi\}=0$.
Hence, the  resulting supersymmetric Hamiltonian is trivial: it coincides with the initial bosonic Hamiltonian.

\subsection*{$\mathcal{N}=4$ supersymmetry}
%
%
For the construction of nontrivial  $\mathcal{N}=4$ supersymmetric system
  we choose the following Ansatz for supercharges
\be\label{Q}
Q_a = f_1(z,\bar z) \psi_a + f_2(z,\bar z)   \psi_a\sum_{b=1}^{\mathcal{N}/2}\psi_b \bpsi^b,\quad
\bQ^a = {\bar f}_1(z,\bar z) \bpsi^a - {\bar f}_2(z,\bar z)   \bpsi^a\sum_{b=1}^{\mathcal{N}/2}\bpsi^b \psi_b \;,
\ee
with
\be
f_1(z,\bar z):=\sqrt{H}{\rm e}^{\imath\Phi_1(z,\bar z)},\qquad f_2=R(z,\bar z){\rm e}^{ \imath(\Phi_1- \Phi_2)},\qquad \bar{R}=R ,\quad \bar\Phi_a(z,\bar z)=\Phi_a(z,\bar z).
\label{12}\ee
Then, we  require  that the supercharges \eqref{Q}  form the $\mathcal{N}= 4$ Poincar\'e  superalgebra \eqref{Poincare},
 which results in the following conditions on the functions involved
\be\label{cond}
\imath \left\{ f_1 , {\bar f}_1 \right\} = f_1 {\bar f}_2 + {\bar f}_1 f_2 \quad\Leftrightarrow\quad
\{\sqrt{H},\Phi_1\}=R\cos \Phi_2,
\ee
with the  Hamiltonian $\mathcal{H}$ acquiring the form
\bea\label{H}
\mathcal{H}&=&   f_1 {\bar f}_1 +  \imath \left\{ f_1 , {\bar f}_1 \right\}\sum_{a=1}^{\mathcal{N}/2}\psi_a \bpsi^a +\frac\imath 2 \left( \left\{ f_1, {\bar f}_2\right\}+
\left\{ f_2, {\bar f}_1\right\} \right)\left( \sum_{a=1}^{\mathcal{N}/2}\psi_a \bpsi^a \right)^2 \\
&=& H+  \{H,\Phi_1\} \psi_a \bpsi^a + A(\sqrt{H},\Phi_{1,2}) \left(\sum_{a=1}^{\mathcal{N}/2}\psi_a \bpsi^a\right)^2,
\eea
with
\begin{align}
&A(\sqrt{H},\Phi_{1,2}):=\frac{\imath}{2} \left( \left\{ f_1, {\bar f}_2\right\}+
\left\{ f_2, {\bar f}_1\right\} \right)= \\ &\left(\{\sqrt{H},\Phi_1 \}\right)^2 -\frac{\{\sqrt{H},\Phi_1 \} \{\sqrt{H},\Phi_2 \}}{\cos^2 \Phi_2}+
\left\{ \{\sqrt{H},\Phi_1 \}, \Phi_1 \right\} \sqrt{H}+ \left\{ \{\sqrt{H},\Phi_1 \}, \sqrt{H} \right\} \tan \Phi_2.
\label{34}\end{align}
Thus, we get the $\mathcal{N}=4$ supersymmetric mechanics parametrized by two arbitrary functions  $\Phi_{1,2}$ .

We can use the Ansatz \eqref{Q} for the construction of  $\mathcal{N}>4$ supersymmetric systems  as well.
However, in that case we get the additional constraints on the functions $f_1, f_2$,
\bea
&\mathcal{N}=6&  : \; \{f_1, {\bar f}_2\} +\{ f_2 ,{\bar f}_1 \}=2\imath f_2  {\bar f}_2\\
&\mathcal{N}=8&  :  \; \{f_1, {\bar f}_2\}  +\{ f_2 ,{\bar f}_1 \}=2\imath f_2   {\bar f}_2,\qquad \{f_2,{\bar f}_2\}=0.
\eea
These constraints, obviously, restrict the functional freedom  existing in $\mathcal{N}=4$ systems.

In the $\mathcal{N}=6$ case this constraint leads to the following restriction
\be
A(\sqrt{H},\Phi_a)+\left(\frac{\{\sqrt{H},\Phi_1 \}}{\cos \Phi_2}\right)^2=0,
\ee
with $A$ given by \eqref{34}.
Hence,  the system has single functional degree of freedom parameterized by $\Phi_1$, as in the $\mathcal{N}=2$ case.

The requirement of $\mathcal{N}=8$ supersymmetry further fixes the value of $\Phi_1$
\be
  {  \{\{\sqrt{H},\Phi_1\},\Phi_1-\Phi_2\}=\{\Phi_1,\Phi_2\}\{\sqrt{H},\Phi_1\}\tan\Phi_2}. \ee
  As a result, we get the    $\mathcal{N}=8$ Hamiltonian  with no  functional freedom.

The evident way to  construct the $\mathcal{N}>4$ systems with wide functional  freedom is to extend the
 supercharges Ansatz by higher $5$- and $7$- fermionic terms .

 \section{$\mathcal{N}=6, 8 $ supersymmetric mechanics}
 In the previous  section we have shown that the supercharges with cubic fermionic terms allow to construct $\mathcal{N}=4$ supersymmetric mechanics with two functional degrees of freedom, $\mathcal{N}=6$ supersymmetric mechanics with single functional degree of freedom,
and $\mathcal{N}=8$ supersymmetric mechanics without any functional freedom.

One can guess that the supercharges  with fifth-order fermionic term could lead to the  $\mathcal{N}=6$ supersymmetric mechanics with three    functional degrees of freedom and to $\mathcal{N}=8$ supersymmetric mechanics with two   functional degrees of freedom.
Furthermore, one can expect that the supercharges  with seventh-order fermionic terms could lead to the  $\mathcal{N}=8$ supersymmetric mechanics with four   functional degrees of freedom and so on.
Let us show that it is indeed the case.

 In order to construct the $\mathcal{N}=6 $ supersymmetric systems with three functional degrees of freedom  we consider
the  following Ansatz for the supercharges
\bea\label{Q8}
Q_a & = &f_1(z,\bar z) \psi_a + f_2(z,\bar z)  \psi_a \left( \sum_{b=1}^{\mathcal{N}/2}\psi_b \bpsi^b \right)\,  + f_3(z,\bar z)  \psi_a\, \left( \sum_{b=1}^{\mathcal{N}/2}\psi_b \bpsi^b \right)^2\; ,\nonumber \\
\bQ^a &=& {\bar f}_1(z,\bar z) \bpsi^a + {\bar f}_2(z,\bar z) \bpsi^a \,\left( \sum_{b=1}^{\mathcal{N}/2}\psi_b \bpsi^b \right)\,
+ {\bar f}_3(z,\bar z)  \bpsi_a\,  \left( \sum_{b=1}^{\mathcal{N}/2}\psi_b \bpsi^b \right)^2  ,
\eea
with $a,b=1,2,3$.

Then, requiring that these functions form $\mathcal{N}=6$ Poincar\'e superalgebra \eqref{Poincare}, we get the following restrictions to the  functions  $f_a$
\be
 f_1 {\bar f}_2 + {\bar f}_1 f_2 - \imath \left\{f_1, {\bar f}_1\right\}=0, \qquad
 2 f_3 {\bar f}_1 + 2 f_2 {\bar f}_2 +2 f_1 {\bar f}_3 - \imath \left\{ f_1 ,{\bar f}_2\right\} +\im \left\{ {\bar f}_1, f_2\right\} =0, \label{82}
\ee

The respective Hamiltonian then reads
\be
\mathcal{H} =  \frac{1}{2}f_1 {\bar f}_1 +\frac{1}{2}( f_1 {\bar f}_2+{\bar f}_1 f_2)  \sum_{a=1}^{\mathcal{N}/2}\psi_a \bpsi^a
+\frac{1}{2}( f_3{\bar f}_1+f_1 {\bar f}_3+f_2 {\bar f}_2)  \left( \sum_{a=1}^{\mathcal{N}/2}\psi_a \bpsi^a \right)^2
 +
\frac{1}{2} (  f_{2}{\bar f}_3+f_3{\bar f}_2) \left( \sum_{a=1}^{\mathcal{N}/2}\psi_a \bpsi^a \right)^3
\label{H6}\ee

Representing $f_a$ in the form
\be
f_1=\sqrt{H}{\rm e}^{\imath\Phi_1},\quad f_2=R_2{\rm e}^{\imath(\Phi_1-\Phi_2)},\quad f_3=R_3{\rm e}^{\imath(\Phi_1-\Phi_2-\Phi_3)},
\ee
and re-writing  in these terms   the conditions \eqref{82},
we conclude that  the functions $\Phi_1,\Phi_2,\Phi_3$ remains unfixed. Therefore, we arrive  at  the family of $\mathcal{N}=6$ supersymmetric mechanics parameterized by three arbitrary real functions.\\

 In order to construct the $\mathcal{N}=  8$ supersymmetric systems with four functional degrees of freedom  we introduce the
   following generalization of the Ansatz \eqref{Q},
\bea\label{Q8}
Q_a & = &f_1(z,\bar z) \psi_a + f_2(z,\bar z)  \psi_a \,  \sum_{b=1}^{\mathcal{N}/2}\psi_b \bpsi^b + f_3(z,\bar z)  \psi_a\,
 \left( \sum_{b=1}^{\mathcal{N}/2}\psi_b \bpsi^b \right)^2
 + f_4 (z,\bar z) \psi_a\, \left( \sum_{b=1}^{\mathcal{N}/2}\psi_b \bpsi^b \right)^3,\nonumber \\
\bQ^a &=& {\bar f}_1(z,\bar z) \bpsi^a + {\bar f}_2(z,\bar z) \bpsi^a \, \sum_{b=1}^{\mathcal{N}/2}\psi_b \bpsi^b
   + {\bar f}_3(z,\bar z)  \bpsi_a\, \left( \sum_{b=1}^{\mathcal{N}/2}\psi_b \bpsi^b \right)^2+ {\bar f}_4 (z,\bar z) \bpsi_a \,\left( \sum_{b=1}^{\mathcal{N}/2}\psi_b \bpsi^b \right)^3 ,
\eea
with $a,b=1,2,3, 4$.

Then, requiring that these functions form $\mathcal{N}=8$ Poincar\'e superalgebra \eqref{Poincare}, we get the following restrictions on the  functions  $f_a$
\bea
& f_1 {\bar f}_2 + {\bar f}_1 f_2 - \imath \left\{f_1, {\bar f}_1\right\}=0, \qquad
  2 f_3 {\bar f}_1 + 2 f_2 {\bar f}_2 +2 f_1 {\bar f}_3 - \imath \left\{ f_1 ,{\bar f}_2\right\} +\imath \left\{ {\bar f}_1, f_2\right\} =0,& \label{44}\\
& 3 f_4 {\bar f}_1+3 f_3 {\bar f}_2 + 3 f_2 {\bar f}_3 +3 f_1 {\bar f}_4 -\imath \left\{f_1,{\bar f}_3\right\}+\imath \left\{{\bar f}_1, f_3\right\} -\imath \left\{ f_2,{\bar f}_2\right\} =0 .&\label{83}
\eea
The respective Hamiltonian then reads
\bea
&\mathcal{H}&= \frac{1}{2}f_1 {\bar f}_1 +\frac{1}{2}( f_1 {\bar f}_2+{\bar f}_1 f_2)\left( \sum_{a=1}^{\mathcal{N}/2}\psi_a \bpsi^a \right)
+\frac{1}{2}( f_3{\bar f}_1+f_1 {\bar f}_3+f_2 {\bar f}_2) \left( \sum_{a=1}^{\mathcal{N}/2}\psi_a \bpsi^a \right)^2 \nonumber \\
 &+&
\frac{1}{2} (f_4{\bar f}_1+f_1 {\bar f}_4+f_{2}{\bar f}_3+f_3{\bar f}_2)\left( \sum_{a=1}^{\mathcal{N}/2}\psi_a \bpsi^a \right)^3
 + \frac{\imath}{8}( \left\{f_1\,,{\bar f}_4 \right\}+\left\{f_2\,,{\bar f}_3 \right\}+\left\{f_3\,,{\bar f}_2 \right\}+\left\{f_4\,,{\bar f}_1 \right\})
\left( \sum_{a=1}^{\mathcal{N}/2}\psi_a \bpsi^a \right)^4
\label{H8}\eea
Let us notice that the restriction \eqref{44} for $\mathcal{N}=8$  system coincides with the restrictions \eqref{82} for $\mathcal{N}=6$ case. While the additional constraint \eqref{83} contains  extra complex function $f_4(z,\bar z)$.
Hence, representing $f_a$ in the form
\be
f_1=\sqrt{H}{\rm e}^{\imath\Phi_1},\quad f_2=R_2{\rm e}^{\imath(\Phi_1-\Phi_2)},\quad f_3=R_3{\rm e}^{\imath(\Phi_1-\Phi_2-\Phi_3)}, \quad f_4=R_4 {\rm e}^{\imath(\Phi_1-\Phi_2-\Phi_3-\Phi_4)},
\ee
we conclude that  the functions $\Phi_1,\ldots,\Phi_4$ remain  unfixed.
Therefore, the $\mathcal{N}=8$ supersymmetric Hamiltonian \eqref{H8} depends on four arbitrary real functions.

{\it So, specifying the formulae given in the {\sl Third and Fours Sections}  to the particular case of Euler top given in  the {\sl Section 2}
by \eqref{fs} we will get its integrable  $\mathcal{N}=2,4,6,8$ supersymmetric extensions}.\\

From the consideration above it is easy to deduce that for the construction of
  $\mathcal{N}=10,12,\ldots 2k$ superextensions of initial Hamiltonian we should choose the following Ansatzes for the supercharges
  \be\label{inf}
Q_a  = f_1(z,\bar z) \psi_a + \sum_{l=1}^{\mathcal{N}/2}f_{l+1}(z,\bar z)  \psi_a \, \left( \sum_{b=1}^{\mathcal{N}/2}\psi_b \bpsi^b \right)^l \, \qquad
\bQ^a = {\bar f}_1(z,\bar z) \bpsi^a + \sum_{l=1}^{\mathcal{N}/2}{\bar f}_{l+1}(z,\bar z) \bpsi^a \,\left( \sum_{b=1}^{\mathcal{N}/2}\psi_b \bpsi^b\right)^l ,
\ee
with $a,b=1,\ldots,\mathcal{N}/2k $.
Then, requiring that they form Poincar\'e superalgebra \eqref{Poincare} we will get the family of  $\mathcal{N}=  2k$ supersymmetric Hamiltonians parameterized by  $k$ arbitrary real functions.

\section{Concluding remarks}
In this paper we formulated the Euler top as a system with    phase space $\mathbf{CP}^1$, i.e. as an one-dimensional system.
Then we proposed the procedure of $\mathcal{N}=2k$ {\sl \'a priori} integrable supersymmetrization of  the generic one-dimensional systems which provides the family of  $\mathcal{N}$-supersymmetric extensions depending on  $\mathcal{N}/2$ arbitrary real functions. Thus, we gave the $\mathcal{N}=2k$ supersymmetric extensions of the Euler top as well.

One may ask whether
  it is possible to construct the family of supersymmetric extensions of the Lagrange and Kowalewski tops (see, e.g.,\cite{perelomov}) which are parameterized by arbitrary functions?

Here we   present some preliminary remarks on this issue.
The phase spaces of Lagrange and Kowalewski tops could be identified with cotangent bundle  of complex projective plane.
This  supermanifold can be equipped  with  three symplectic (and complex) structures,
 parameterized by the coordinates $u^A=(z ,\pi )$,
 \bea
 &\omega_1 = d\pi \wedge dz  +d\bar\pi \wedge d\bar z 
 \qquad
 \omega_2= \imath d\pi \wedge dz  -\imath d\bar\pi \wedge d\bar z ,
 &\label{omega2}\\
 &\omega_3=   \imath\frac{\partial^2{\tilde K}}{\partial u^A\partial\bar u^B}du^A\wedge d\bar u^B,\quad {\tilde K}=
 K(z,\bar z)+F(g^{-1}\bar\pi  \pi ),&\label{omega3}
 \eea
  where $K(z,\bar z)$ and $g( z,\bar z)$ are given by \eqref{fskah} and\eqref{fs} respectively, while $F(x)$ is the real function obeing condition $F'(0)\neq 0$.
Within appropriate choice   of the function $F(x)$  these symplectic structures provide  the manifold $T^*\mathbf{CP}^1$ with hyper-K\"ahler structure \cite{perelomovPR}.
Formulating Lagrange and Kowalewski tops in terms of symplectic structures \eqref{omega2}, we can try to construct their
 conventional ${\mathcal N}=2,4$ supersymmetric extensions, extending these simplectic structure by fermionic variables associated with
$dz$.
%
However, we are expecting that using the symplectic structure \eqref{omega3} will be more useful for the construction of the supersymmetric extensions of Lagrange and Kowalewski tops.

\acknowledgements

This work was supported by the Russian Foundation of Basic Research grant 20-52-12003 (S.K., A.N.) and by the
Armenian Science Committee projects 20RF-023, 21AG-1C062 (E.Kh., A.N.) and 21AA-1C001(E.Kh.). The work
of E.Kh. was completed within the Regional Doctoral Program on Theoretical and Experimental Particle Physics
Program sponsored by VolkswagenStiftung.

\end{document}